\documentclass[journal]{vgtc} 

\usepackage{cite}
\usepackage[bottom]{footmisc}
\usepackage{mathptmx}
\usepackage{graphicx}
\usepackage{siunitx}
\usepackage{times}
\usepackage[T1]{fontenc}
\usepackage{enumitem}
\usepackage[table]{xcolor} 
\usepackage{wrapfig}
\usepackage{multirow}
\usepackage{adjustbox}
\usepackage{mwe}
\usepackage{graphbox}
\usepackage{booktabs}
\usepackage{dirtytalk}
\usepackage{soul}
\usepackage{mdframed}
\usepackage{tabularx}
\usepackage{xcolor}
\usepackage{enumitem}
\definecolor{CindySalmon}{RGB}{232, 125, 114}
\definecolor{greenB}{RGB}{77, 175, 74}
\definecolor{purpleF}{RGB}{152,78,163}

\usepackage[bookmarks,backref=true,linkcolor=black]{hyperref} 
\hypersetup{
 pdfauthor = {},
 pdftitle = {},
 pdfsubject = {},
 pdfkeywords = {},
 colorlinks=true,
 linkcolor= black,
 citecolor= black,
 pageanchor=true,
 urlcolor = black,
 plainpages = false,
 linktocpage
}

\onlineid{1062} 
\vgtccategory{Theoretical and Empirical} 
\vgtcinsertpkg 

\title{Dispersion vs Disparity: Hiding Variability Can Encourage Stereotyping When Visualizing Social Outcomes}

\author{Eli Holder and Cindy Xiong}
\authorfooter{ 
\item
Eli Holder is with 3iap
\newline
E-mail: eli@3iap.com
\item
Cindy Xiong is with UMass Amherst
\newline
E-mail: cindy.xiong@cs.umass.edu
}

\shortauthortitle{Holder \MakeLowercase{\textit{et al.}}: Dispersion vs Disparity}

\abstract{Visualization research often focuses on perceptual accuracy or helping readers interpret key messages. However, we know very little about how chart designs might influence readers' perceptions of the people behind the data. Specifically, could designs interact with readers' social cognitive biases in ways that perpetuate harmful stereotypes? For example, when analyzing social inequality, bar charts are a popular choice to present outcome disparities between race, gender, or other groups. But bar charts may encourage deficit thinking, the perception that outcome disparities are caused by groups’ personal strengths or deficiencies, rather than external factors. These faulty personal attributions can then reinforce stereotypes about the groups being visualized. We conducted four experiments examining design choices that influence attribution biases (and therefore deficit thinking). Crowdworkers viewed visualizations depicting social outcomes that either mask variability in data, such as bar charts or dot plots, or emphasize variability in data, such as jitter plots or prediction intervals. They reported their agreement with both personal and external explanations for the visualized disparities. Overall, when participants saw visualizations that hide within-group variability, they agreed more with personal explanations. When they saw visualizations that emphasize within-group variability, they agreed less with personal explanations. These results demonstrate that data visualizations about social inequity can be misinterpreted in harmful ways and lead to stereotyping. Design choices can influence these biases: Hiding variability tends to increase stereotyping while emphasizing variability reduces it.}


\begin{document}
\maketitle
\section{Introduction}
Understanding (and addressing) social inequity often begins with quantifying outcome disparities. For example, in Data Feminism, D'Ignazio and Klein tell the story of Christine Darden who blazed trails in aerodynamics and more equitable career outcomes at NASA \cite{d2020data}. The latter achievement began with a bar chart, showing disparate promotion rates between white men and women of color. Upon seeing this chart, NASA leadership was “shocked at the disparity.” The analysis created awareness about the systemic nature of the disparities, clearing an early obstacle for Darden and others to take on more significant roles within the organization.


Practitioners often use bar charts to communicate inter-group differences \cite{szafir2018good}, where each marker encodes the average outcome for group members (via position and length). This technique, while conventional and accurate \cite{cleveland1986experiment, xiong2019biased, mccoleman2021rethinking}, can lead to systematic misinterpretation of data \cite{streit2014bar, kale2018hypothetical, andrade2011excessive, lan2022negative}. For example, 1 in 5 people think that the tip of each bar represents the outer limit of the group data, rather than the average value \cite{kerns2021bar, kerns2021two}, leading to the phenomenon where people tend to think data values that fall within the bar are more likely than points outside the bar \cite{newman2012bar}. The presentation of mean differences in bar charts can influence how they are assessed \cite{srinivasan2018s, nguyen2020exploring}.

When visualizing social outcomes -- especially between dominant and minoritized groups -- these conventional approaches have an additional risk. They encourage \emph{deficit thinking} - a perspective that social groups with worse outcomes are somehow personally deficient \cite{davis2019deficit}. That is, through a process of rationalizing outcome disparities, a number of social cognitive biases conspire to reinforce stereotypical beliefs about the groups in focus \cite{stangor2014social}.



Conventional data design choices reinforce this deficit framing. Charts that emphasize differences \emph{between} groups (e.g. as point estimates of group averages) and downplay differences \emph{within} groups (i.e. intra-group outcome variability) can create an exaggerated sense of certainty \cite{manski2020lure, hullman2019authors} about how well a summary statistics represents the whole group;  essentially the group average acts as a quantitative stereotype (e.g. seeing that Group B has worse \emph{average} outcomes than other groups, viewers can mistakenly conclude that the average in the graph applies to all members of Group B and, therefore, everyone in Group B must have worse outcomes than all other people). The false conclusions of group homogeneity (i.e. low intra-group outcome variance) then support another: that the worse outcomes are \emph{caused by} intrinsic personal attributes, therefore the groups with the worst outcomes are personally deficient \cite{stangor2014biases, stangor2014social, xiong2019illusion}. 

Research so far focuses on chart selection to maximize perceptual accuracy or increase cognitive affordances \cite{xiong2021visual, mccoleman2021rethinking}. But we know very little about how visualization designs can influence viewers' social perceptions of the people depicted in data. With visualization becoming more ubiquitous for communication of social issues (e.g., in data-driven journalism), it is more critical than ever to consider how visualizations influence perceptions about the people being visualized.

\vspace{1mm}
\noindent \textbf{Contributions:} We contribute four controlled experiments demonstrating that visualization readers are subject to attribution bias, and this bias can be exacerbated or mitigated by visualization design choices. Our results suggest that emphasizing within-group outcome variability can reduce readers' tendencies toward personal attributions (and therefore stereotyping).




\section{Related Work}

\subsection{Deficit Thinking and Stereotyping}
Stereotypes are over-generalizations about a group of people \cite{oakes1994stereotyping}. 
Stereotypes are harmful when we let them determine our expectations about individual members of a group \cite{stangor2014social}. 
Stereotyping is facilitated by perceptions of group homogeneity; when we overestimate the similarity of people within a group, it’s easier to apply stereotypes to individual members of the group \cite{stangor2014social}. Unfortunately, we’re predisposed toward overestimating group homogeneity for other groups than our own \cite{park1982perception, linville1980polarized}. 
Our faulty judgments about other people reinforce stereotypes. For example, we often attribute others’ successes or failures to personal qualities, even when the outcomes are obviously outside their control \cite{lerner1965evaluation}.
Compounding this, when we observe something negative about a person from another group, we tend to associate similar negative attributes with other members of the group (reinforcing stereotypes)\cite{hamill1980insensitivity, allison1985group}. 
These biases help explain the risks of deficit-framing. By emphasizing between-group differences 1) perceptions of within-group homogeneity become exaggerated, 2) the lower-outcome groups (subconsciously) take the blame, and 3) the faulty personal attributions are then amplified to the entire group.
Prejudicial tendencies can be overcome. The more exposure we have to people from other groups, the less likely we are to stereotype them (exposure helps us appreciate group heterogeneity)\cite{pettigrew2006meta, stangor2014reducing}. 

Practitioners have argued that certain charts encourage deficit thinking. By emphasizing direct comparisons between groups, they create the (false) impression that groups with the worst outcomes (often marginalized groups) are personally deficient relative to groups with the best outcomes (often majority groups) \cite{schwabish2021no, blakely2021presenting}.
Deficit thinking encourages ``victim blaming;'' it favors explanations that hold group members personally responsible for outcomes (e.g., “It’s because of who they are”), as opposed to explanations related to external causes (e.g., “It’s because of systemic racism”) \cite{davis2019deficit}. This perspective reinforces harmful stereotypes, setting lower expectations for minoritized groups that become self-fulfilling prophecies \cite{stangor2014social}.
This bias can occur unconsciously and is difficult to inhibit \cite{samson2005seeing, stangor2014biases, kahneman2011thinking}, locking readers into a tunnel vision that leaves more complex (external), but widespread, systemic problems unconsidered and unaddressed \cite{davis2019deficit}.

\subsection{Communicating Uncertainty}
Showing only summary statistics can create a false impression of homogeneity, misleading viewers to think that groups of data have low within-group variability \cite{andrade2011excessive, soyer2012illusion}. Despite acknowledging the value of depicting uncertainty, visualization authors often omit it due to limited resources, concerns for making errors, or worries about reader understanding \cite{hullman2019authors}. While it's true that readers can misinterpret uncertainty \cite{kahneman1982judgment, padilla2017effects}, certain visualization designs can enhance their understanding \cite{xiong2022investigating, hofman2020visualizing, kale2020visual, kay2016ish}. For example, gradient and violin plots outperform bar charts (even with error bars) \cite{correll2014error}, Hypothetical Outcome Plots (HOPs) outperform violins \cite{hullman2015hypothetical}. Even showing predicted hurricane paths as ensemble plots leads to better understanding compared to monolithic visualizations \cite{ruginski2016non, cox2013visualizing}. 

Showing uncertainty in data can introduce other revealing forms of cognitive bias.
Hofman et al. show that when viewers underestimate outcome variability (by mistaking 95\% CIs for 95\% PIs), they overestimate the effects of certain treatments \cite{hofman2020visualizing}. And Xiong et al. show a link between data granularity and perceived causation \cite{xiong2019illusion}; more granular (less aggregated) charts minimize mistaking correlation for causation. These effects might be explained by visualization readers' over-sensitivity to co-occurrences and a tendency to generalize based on limited data \cite{kahneman1982judgment}. By disclosing more variability in data, visualization authors provide readers more counterexamples, which can mitigate cognitive biases and encourage critical thinking \cite{matute2015illusions, blanco2013interactive}. 

Visualization designs can clarify or obscure the shape of the underlying data distribution. If stereotypes stem from underestimating the dispersion of within-group outcomes (homogeneity), then designs that hide variability (e.g. bar charts) might encourage stereotyping relative to designs that emphasize variability (e.g. jitter plots). 


\section{Study Motivation and Overview}
We conducted four experiments to understand how data visualization can influence viewers' explanations of inter-group disparities in social outcomes, specifically their tendencies toward personal attributions (e.g., ``their data looks like this because of who they are'') or external attributions (e.g., ''their data looks like this because of reasons outside their control'').
Stimuli, survey, data, and analysis script are available here: \url{https://osf.io/r4cwa/?view_only=66f42a49c9844fd7a3ce03941d9a1f20}. Based on existing work, we make the following hypotheses:
\newline
\vspace{-2mm}


\noindent \textbf{Hypothesis No.1.} When interpreting visualizations depicting social outcome disparities, viewers' reported beliefs may reflect causal explanations not supported by the visualized data, but consistent with documented social cognitive biases (e.g., attribution errors).
\newline
\vspace{-2mm}


\noindent \textbf{Hypothesis No.2.} Visualizations that \emph{mask} within-group outcome variability (e.g., bar charts, dot plots, sans error bars) may encourage stronger personal attributions compared to charts that \emph{emphasize} within-group variability (e.g., jitter plots, prediction intervals).
\newline
\vspace{-2mm}


\noindent \textbf{Hypothesis No.3.} The type of uncertainty visualized may influence viewers' attributions. Visualizing inferential uncertainty (e.g., confidence intervals) may encourage stronger personal attributions compared to visualizing outcome uncertainty (e.g., prediction intervals).
\newline
\vspace{-2mm}

%

\noindent \textbf{Hypothesis No.4.} For unit charts (e.g., jitter plots), the number of people (data points) shown per group may influence attribution.
\newline
\vspace{-2mm}

In each experiment, we presented participants with visualizations depicting disparities in a particular outcome (e.g., income) across 3-4 hypothetical groups of people. Users reported their agreement with statements based on either personal or external causes for the visualized disparities. Each experiment compares a different set of visualizations, chosen to test the above hypothesis. 


\section{Experiment 1: Bar and Jitter Plots}
When interpreting data about people, viewers can form mental explanations for why the data follows a certain pattern, even when the data offers no such explanation. The explanation can be attributed to personal or external causes. In Experiment 1, we investigate whether hiding or disclosing within-group variability impacts viewers' agreement with personal or external explanations of social outcome data. 

\begin{figure}[h!]
\centering
 \includegraphics[width = \columnwidth]{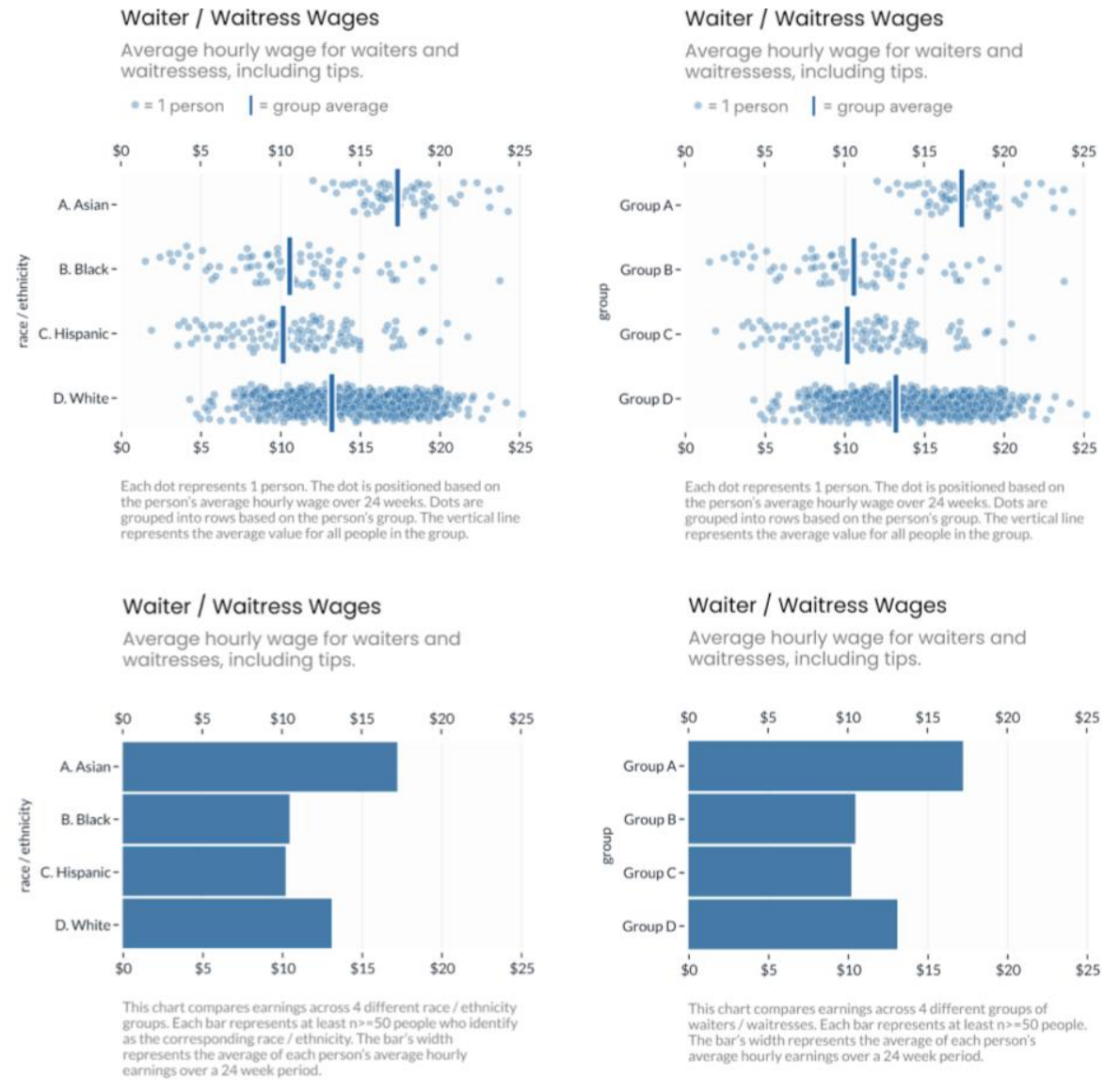}
 \caption{Bar charts and jitter plots from Experiment 1, labeled with ambiguous letters (left) or explicit racial information for each group (right).}
 \label{Exp1Stimuli}
\vspace{-1em}
\end{figure}

\subsection{Experiment 1: Stimuli and Design}
We generated two types of visualizations that either hide or disclose within-group variability: bar charts or jitter plots, as shown in Figure \ref{Exp1Stimuli}. The bar charts only show differences between group means and convey nothing about within-group variability. The jitter plots convey variability as each jittered dot corresponds to the data of one individual. We added mean markers so the dot plots also explicitly convey between-group differences, to be consistent with the bar charts.

Additionally, due to the social desirability bias \cite{stark2019impact, sears2005over}, participants might refrain from expressing potential prejudices when they view charts labeled with explicit racial information, in hopes of being viewed more favorably by the experimenters \cite{goffman1978presentation}. To account for potential effects of experimenter demand and social desirability bias with regards to stereotyping \cite{grimm2010social, hopkins2009partisan}, we showed participants the charts either labeled with ambiguous letters (A, B, C, D) or with explicit race/ethnic group information (Asian, Black, Hispanic, White). 

We generated the visualizations referencing data from the US Bureau of Labor Statistics \cite{waiters2021waitresses} depicting wages for waitstaff. We adjusted the data to match proportional overall pay disparities between race/ethnicity groups \cite{economic2021table_wage}. For the dot plots, the values were normally, randomly distributed. We selected waitstaff wages because it's a realistic topic that might reasonably elicit either personal or external explanations of the disparities. (More on this in Experiment 2, where we additionally investigate the effect of topic.)

\subsection{Experiment 1: Procedure}
Participants saw a single page with a single chart and three sections. 
The chart was randomly assigned to be one of the four from Figure \ref{Exp1Stimuli}. 
The first section checked for comprehension and attention. See “Participants” for notes on exclusion criteria. 
In the second section, participants rated how much they agreed with external (e.g., ``Based on the graph, Group A likely works in a more expensive restaurant than Group D.'') or personal (e.g., ``Based on the graph, Group A likely works harder than Group D.'') explanations of the disparities.
To keep the length of the experiment manageable and avoid the fatigue effect, participants were randomly assigned 5 questions soliciting agreement with personal attributions and 5 questions soliciting agreement with external attributions, from a pool of 33 similar questions (18 personal, 15 external). 
The questions were presented in random order. Participants responded by moving a slider initialized at 50 (neither agree nor disagree), ranging from 0 (strong disagree) to 100 (strong agree).
The third section asked participants to explain the visualized disparities in their own words and to self-identify with a race/ethnicity and a political party. Race and political affiliation were asked to control for possible in-group or out-group biases \cite{weiner2011attributional, osborne2015latent, hopkins2009partisan, stangor2014biases}.


\subsection{Experiment 1: Participants and Exclusions}
We recruited participants via Amazon's Mechanical Turk and pre-screened for those who are located in the U.S., have 98\%+ MTurk approval rating, and did not participate in any of the pilot tests. To filter out spammers, we also filtered for participants that correctly answered two attention checks (e.g., 2+2=?), did not put nonsensical responses in the free-response questions, and correctly answered a simple set of numeracy questions (e.g., identify the highest earning group).

Power analysis based on an omnibus effect suggested that 280 participants would be sufficient to see the expected effect size on chart design from previous pilot studies (for 80\% power at 5\% significance). To account for losing participants to the above criteria, (531 participants were recruited across 2 sessions to account for potential exclusions, randomly assigned to view either the jitter or the bar plot. After applying the exclusion criteria, we ended up with 329 participants 152 saw the jitter plot, and 177 saw the bar chart), with the group type (chart labeled with letters or race) counterbalanced. 
Since we collected data from more participants than indicated by the power analysis, we conducted a post-hoc power analysis to assess its impact on our results.

\begin{figure*}[h!]
\centering
 \includegraphics[width = \linewidth]{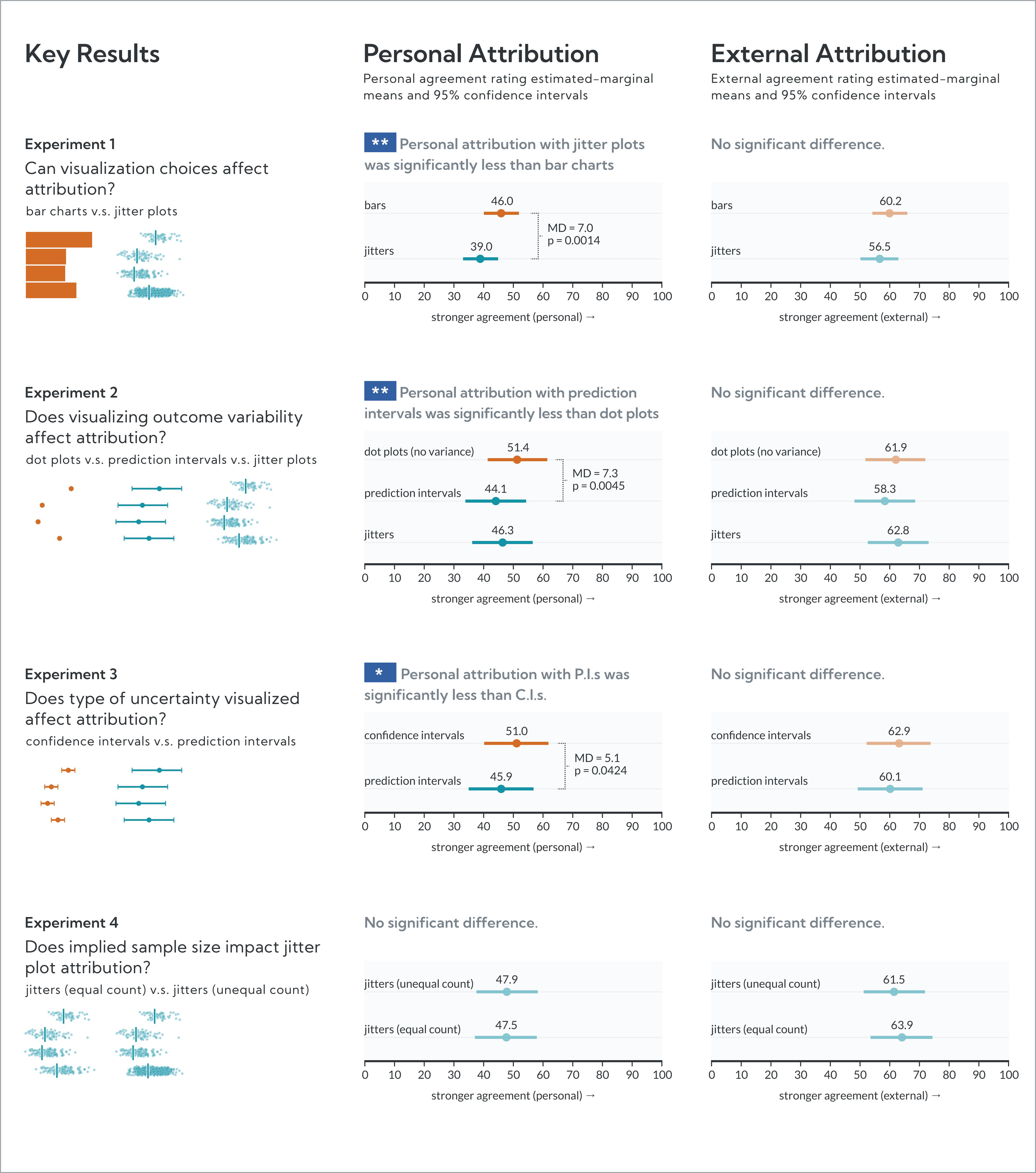}
 \caption{Key results comparing agreement ratings with personal and external attributions for all experiments. Orange indicates designs with low variability salience, teal indicates designs with high variability salience. Asterisks indicate significant Tukey-adjusted pairwise differences. Results and confidence intervals are based on estimated marginal means (interval overlap should not be used for inference).}
 \label{fig:Overall_Results}
\end{figure*}

\subsection{Experiment 1: Results}
We constructed a mixed-effect linear model to fit participants' agreement ratings \cite{bates2005fitting} with question type (personal or external), chart type (bar or jitter plot), and their interactions. 
We additionally included participant race, political orientation, and chart labels (ambiguous letters or explicitly racial groups) as fixed effects.
We used a random intercept term accounting for individual differences and question index as random effects. 

\begin{figure}[!h]
\centering
 \includegraphics[width = \columnwidth]{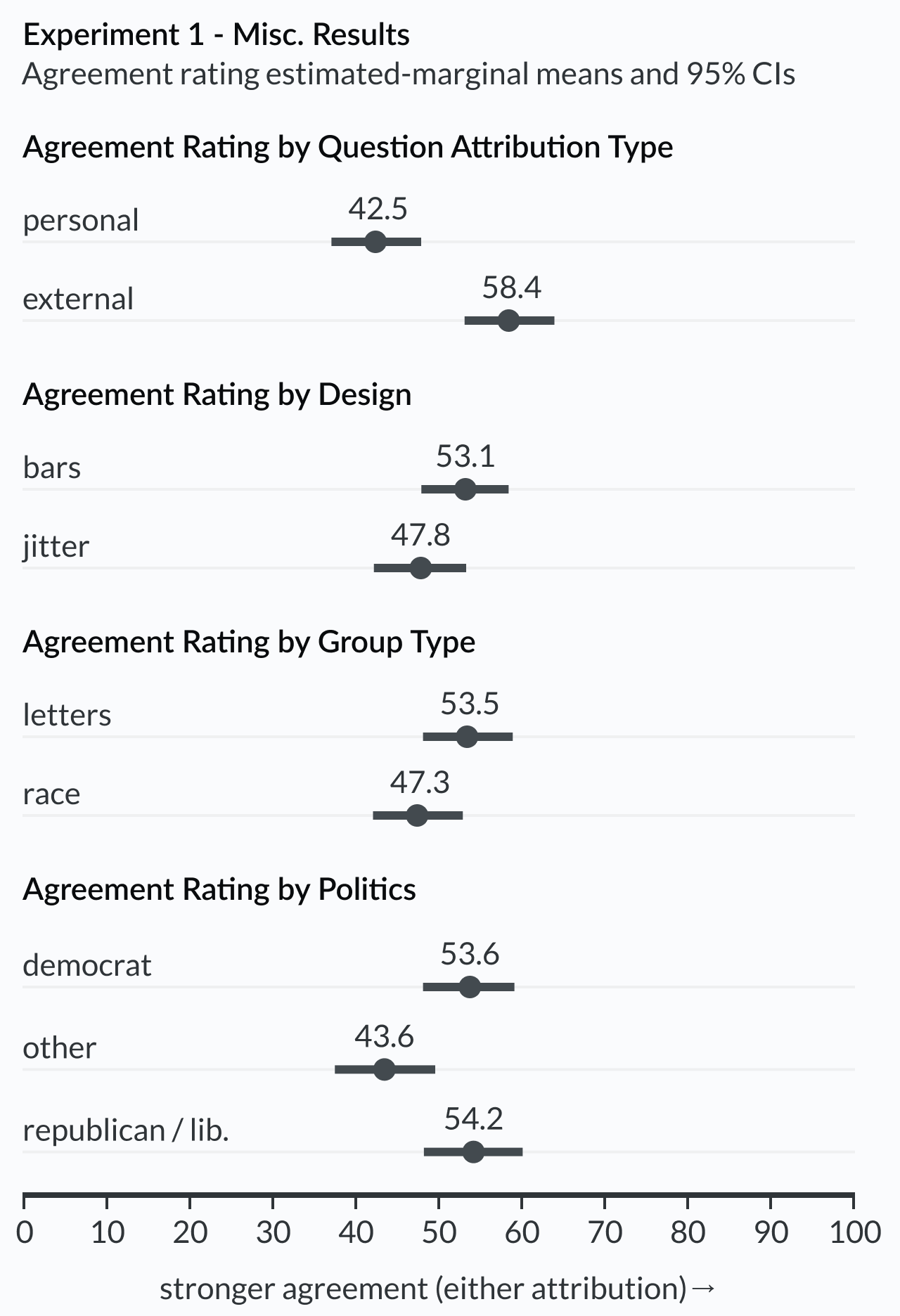}
 \caption{Experiment 1 agreement rating estimated-marginal means and 95\% CIs across factors for both personal and external attributions. First: Ratings by question attribution type. Second: Ratings for chart type. Third: Ratings for charts labeled with letters or racial information. Fourth: Ratings for participants across political orientations. Results and CIs are based on estimated marginal means (interval overlap should not be used for inference).}
 \label{Fig:Exp1_QType_GroupType}
\end{figure}


We found a significant main effect of question type (F = 35.99(1, 30.59), p < 0.001, $\eta^2$ = 0.54), such that participants agreed more with external explanations of the data rather than personal explanations (MD = 15.9, p < 0.0001, d = 0.724), as shown in the first panel of Figure \ref{Fig:Exp1_QType_GroupType}. There is also a main effect of chart type (F = 6.71(1, 322.09), p = 0.010, $\eta^2$ = 0.02). People agreed slightly more with both personal and external explanations of data when they viewed bar charts (MD = 5.37, p = 0.0096, d = 0.245).
Post-hoc power analysis suggests 73.4\% power with our sample size and effect size (alpha = 0.05).

There is a significant interaction between chart type and question type (F = 4.83(1, 2951.64), p = 0.028), which is driven by participants agreeing more with \textit{personal} explanations of data patterns when viewing bar charts compared to dot plots, according to post-hoc pair-wise comparison with Tukey's adjustment (MD = 7.06, p = 0.0014, d = 0.322). There is no significant difference between participant agreement ratings on external explanations between bar charts and jitter plots (MD = 3.68, p = 0.096). 

\subsubsection{Effects of Group Type}
As shown in Figure \ref{Fig:Exp1_QType_GroupType}, there appears to be a significant effect of chart label (F(1, 321.99) = 8.97, p = 0.0030, $\eta^2$ = 0.03) such that participants agreed less with both personal and external explanations of data when the charts are explicitly labeled with race, compared to when the charts are labeled with ambiguous letters.

\subsubsection{Effects of Participant Demographics}
We analyzed potential effects of co-variate demographic factors on our key dependent variable. 
Overall, participants' race was not a significant factor affecting agreement ratings (F(4, 320.06) = 0.93, p = 0.45). 
Political orientation was a significant predictor for agreement ratings (F(2,319.93) = 7.84, p < 0.0005, $\eta^2$ = 0.05).
It's important to note that we controlled for the potential effects of politics by treating it as a fixed-effect predictor in the model,
but the current experiment does not control for these effects in the design, so the results are purely descriptive.
This experiment was not designed to identify the potential effects of political orientation on the tendency to make attribution errors. We discuss future opportunities to do this optimally in Section \ref{limitsFuture}.

\subsection{Experiment 1: Discussion}

Consistent with Hypothesis No.1, participants agreed with causal explanations of the outcomes that were not supported by the visualized data. While they expressed stronger agreement for external attributions, 38\% of participants agreed with personal attributions (i.e., their average personal attribution response was >= 51).
More critically, personal attribution agreement was stronger when participants viewed bar charts, compared to jitter plots. This supports Hypothesis No.2, suggesting that visualization design  (specifically, those emphasizing within-group variability) can influence attribution biases.
The type of group visualized also influenced attribution agreement (both external and personal). Participants expressed stronger attribution agreement when the charts visualized groups labeled with ambiguous letters, compared to charts with explicitly racial groups. This may suggest that participants intuitively recognized that attributions based on race are unreasonable. Or, as previously noted, it could suggest that the presence of race triggered social-desirability biases \cite{stark2019impact, sears2005over, goffman1978presentation}, causing participants to artificially suppress attribution tendencies. To address this potential bias, we remove the racial group labels condition from subsequent experiments.


\begin{figure}[h!]
\centering
 \includegraphics[width = \columnwidth]{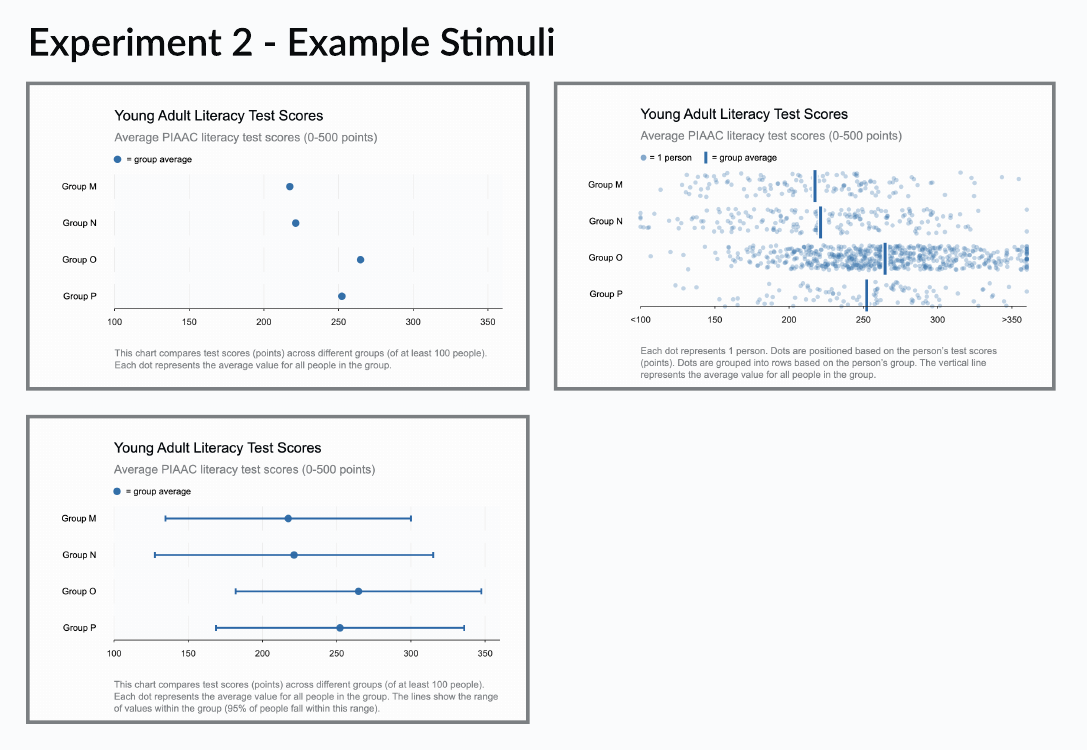}
 \caption{Dot plot, prediction interval plot, and jitter plot used in Experiment 2. In the experiment, the charts depicted one of three topics: household income, literacy scores, and life expectancy. This figure shows the charts depicting the literacy score topic as examples.}
 \label{Exp2Visualizations}
\vspace{-1em}
\end{figure}

\section{Experiment 2: No Variance, Interval, and Jitter}
In Experiment 2, we further test whether showing within-group variability can mitigate attribution bias. 
Considering that the bar charts and the jitter plots used in Experiment 1 differed both in the amount of within-group variability they show and the visual encoding marks (bars vs. dots), we control for the effect of visual encoding marks in Experiment 2 by testing dot plots (with no error bars) against jitter plots. 
Dot plots are similar to bar charts as they both use position to encode the mean group values, but dot plots provide a symmetrical representation of the mean values, without risking the within-the-bar bias \cite{newman2012bar}. 
This change enables us to compare the effect of showing within-group variability without worrying about the confounding factor of visual encoding marks. 
We also introduced a third visualization type, dot plot with prediction intervals, to test whether \textit{how} within-group variability is displayed has an effect. This also enables us to isolate unit-encoding to observe its effect. The three charts are shown in Figure \ref{Exp2Visualizations}. 
Furthermore, we included three topics (household income, literacy scores, and life expectancy) to test whether the effects are robust across topics.


\subsection{Experiment 2: Participants}
We followed the same procedure of recruiting participants and a similar protocol to filter for spammers and invalid answers as to that in Experiment 1. Additionally, with the growing number of invalid responses researchers are collecting from MTurk, we added a pre-filter for VPN-suspicious IPs following recommended best practices from social science research \cite{kennedy2020shape, agley2022quality}. 
Power analysis based on an omnibus effect suggested 375 participants would achieve 80\% power to see the expected effect size on chart design from previous pilot studies at an alpha level of 0.05.
After applying exclusion criteria, we are left with 434 participants (162 who saw the no variance dot plot, 140 who saw the prediction interval plot, and 132 who saw the jitter plot).

\subsection{Experiment 2: Stimuli and Design}
We used a between-subject design where each participant was randomly assigned to view one chart depicting one topic. 
The jitter plot is similar to the one used in Experiment 1, but we accounted for the difference in within-group sample size by making the number of data points in each group identical, with n = 300 per group. 

Based on the time it took for participants to complete Experiment 1, we further reduced the number of questions asked of participants, for both external and personal explanations, from five to two each, to make the experiment length more manageable. The questions, similar to the setup in Experiment 1, were chosen from a set of external and personal attribution questions. The questions were presented in random order. Participants responded by moving a slider initialized at 50 (neither agree nor disagree), ranging from 0 (strong disagree) to 100 (strong agree). 

\subsection{Experiment 2: Results}
We again constructed a mixed-effect linear model to fit participants' ratings of agreement \cite{bates2005fitting} with question type (personal or external), chart type (no-variance dot plot, prediction interval plot, and jitter plot), and their interactions. 
We additionally included chart topics (literacy scores, household income, and life expectancy), participant race, political orientation, and participant gender as fixed effects.
We used a random intercept term accounting for individual differences and question index as random effects. 

We found a significant main effect of chart design (F(2, 421.07) = 4.10, p = 0.017, $\eta^2$ = 0.02).
Post-hoc comparison with Tukey's adjustment reveals that this effect is driven by the difference between the no-variance dot plot and the prediction interval; participants reported stronger agreement for both personal and external attribution when looking at the no-variance dot plot (MD = 5.45, p = 0.013, d = 0.245).

We found a significant main effect of question type (F(2, 5.96) = 6.18, p = 0.048, $\eta^2$ = 0.51). 
Participants more highly agreed with external compared to personal attributions (MD = 13.7, p = 0.0478, d = 0.616).
Post-hoc power analysis indicates that with our current sample size we have 84.4\% power to detect our given effect for design, and 100\% power to detect that for question type at an alpha level of 0.05.

We found a trending significant interaction between chart type and question type (F(2, 1293.94) = 2.72, p = 0.066). 
Post-hoc comparison with Tukey's adjustment reveals that this effect is driven by significant differences in agreement ratings for the personal explanations for data.
Specifically, participants agreed most  with personal explanations for the data when it's visualized as a no-variance dot plot, compared to a prediction interval plot (MD = 7.30, p = 0.0045, d = 0.328), and a jitter plot (MD = 5.08, p = 0.075, d = 0.228), as shown in Figure \ref{fig:Overall_Results}. 

\begin{figure}[h!]
\centering
   \includegraphics[width = \columnwidth]{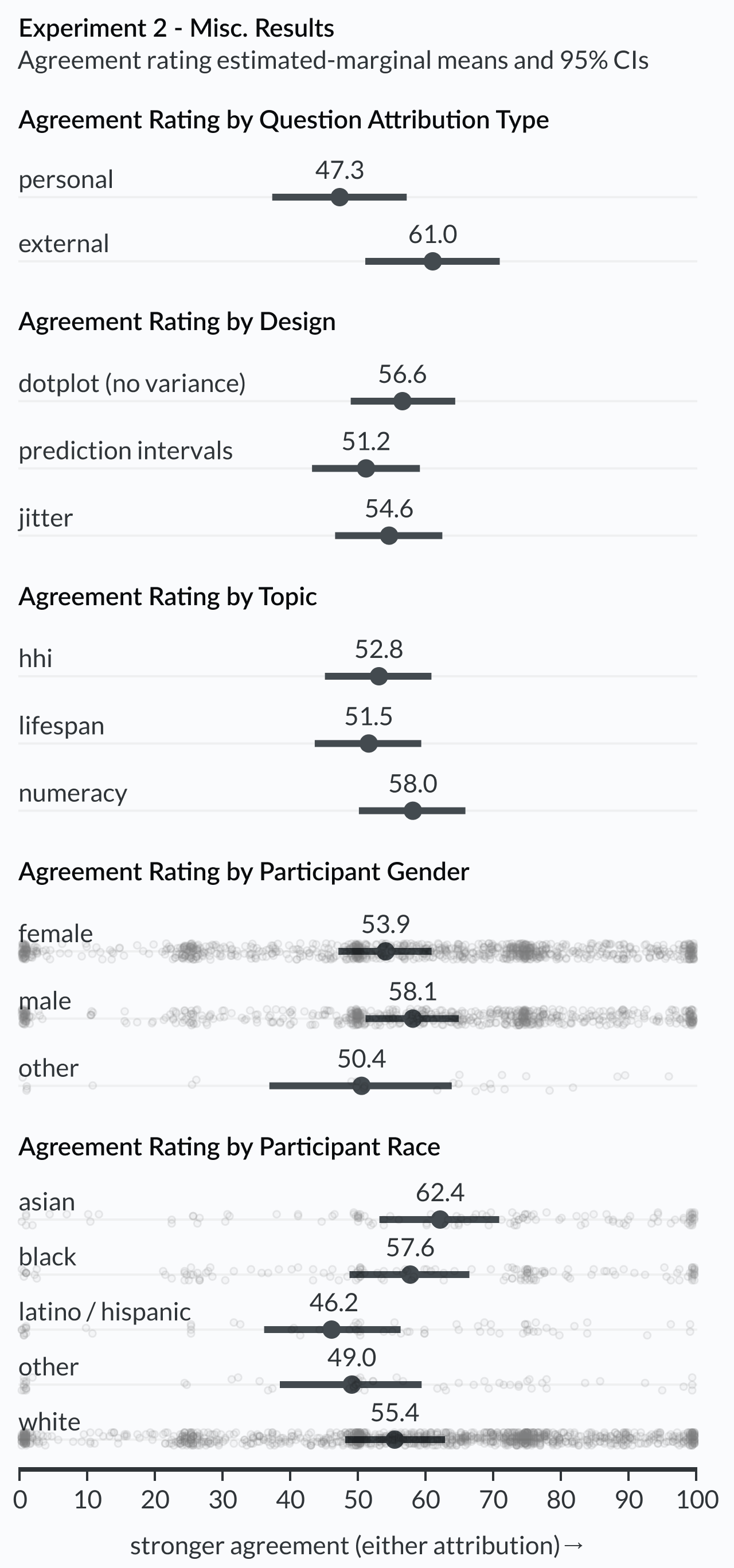}
 \caption{Experiment 2 agreement rating estimated-marginal means and 95\% CIs across factors for both personal and external attributions. First: Ratings by question attribution type. Second: Ratings for chart type. Third: Ratings by topic. Fourth: Ratings by participants' self-identified gender. Fifth: Ratings by participants' self-identified race/ethnicity. Gender and race charts also show jitter plots of individual responses. Results and CIs are based on estimated marginal means (interval overlap should not be used for inference).}
 \label{Exp2TopicGenderRace}
\vspace{-1em}
\end{figure}

\subsubsection{Effects of Topic and Participant Demographics}
We created visualizations using datasets from three topics to increase the external validity and the generalizability of our results.
Though not an experimental variable of interest, we did find a small but significant effect of topic (F(2, 421.13) = 5.95, p = 0.0028, $\eta^2$ = 0.03).
As shown in the third panel of Figure \ref{Exp2TopicGenderRace}, participants were more likely to agree with explanations of outcome disparity overall when viewing charts depicting the numeracy topic. 

As shown in the bottom panels of Figure \ref{Exp2TopicGenderRace}, we observed a significant effect of gender on agreement ratings (F(2, 421.12) = 3.64, p = 0.027, $\eta^2$ = 0.02), as well as a significant effect of race (F(4, 421.12) = 3.513, p = 0.0078, $\eta^2$ = 0.03). But this time we observed no significant effect of political orientation (F(2, 421.26) = 2.27, p = 0.105).

\subsection{Experiment 2: Discussion}
Consistent with Experiment 1 (and Hypothesis 1), participants agreed with causal explanations of the outcomes that were not supported by the visualized data. While they expressed stronger agreement for external attributions, many agreed with personal attributions as well.
We again found that personal attribution agreement was influenced by the presence of outcome variability in the charts; personal attribution agreement was significantly stronger for dot plots (sans error bars) compared to prediction intervals. Combined with results from Experiment 1, this suggests that showing within-group outcome variability reduces personal attribution bias, supporting Hypothesis No.2. Although this result seems to generalize across topics and demographics, the present experiment was not designed to observe nuanced individual differences. We discuss future opportunities to further investigate this in Section \ref{limitsFuture}.

\section{Experiment 3: Confidence and Prediction Intervals} 
In Experiment 2, both jitter plots and prediction intervals communicate within-group outcome uncertainty. Confidence intervals, however, are a perhaps more common type of uncertainty visualization used in practice. Unlike prediction intervals, which represent outcome uncertainty (i.e., variation of outcomes around the mean, using standard deviation), confidence intervals represent inferential uncertainty (i.e., uncertainty in the estimate of a mean, using standard error).
Hofman et al. show that confidence intervals lead viewers to overestimate effect sizes, compared to prediction intervals \cite{hofman2020visualizing}. They also suggest that this might contribute to stereotyping when visualizing inter-group social outcomes. This implies that personal attributions could be influenced not just by the \emph{presence} of uncertainty within a visualization, but by the \emph{type} of uncertainty.

The difference between inferential and outcome uncertainty has a salient visual consequence: Confidence intervals are relatively short compared to prediction intervals. And confidence intervals can be made arbitrarily small with increased sample sizes. So prediction intervals might show long, overlapping ranges, while confidence intervals of the same data are shorter and visually separate. This becomes problematic when considering practitioners' inconsistency with error bars \cite{krzywinski2013errorbars} and (lay) readers' tendencies to misread them \cite{belia2005researchers, hofman2020visualizing, correll2014error}. 
We suspect that in Experiments 1 and 2, the differences in personal attribution are related to outcome range salience. Jitter plots and prediction intervals convey a wide range of outcomes and seeing them overlap reminds readers that people across groups often experience similar outcomes, undercutting group membership as a cause for disparities and therefore tendencies toward personal attribution. If this is the case, and if lay audiences tend to misread confidence and prediction intervals as if they're simply ranges, then shorter, non-overlapping intervals would elicit more personal attribution than longer, overlapping intervals.

To shed light on whether the type of uncertainty (and consequent interval sizes) visualized has an effect on attribution biases, in Experiment 3, we compare the effect of visualizing within-group uncertainty using prediction versus confidence intervals.


\begin{figure}[h!]
\centering
 \includegraphics[width = \columnwidth]{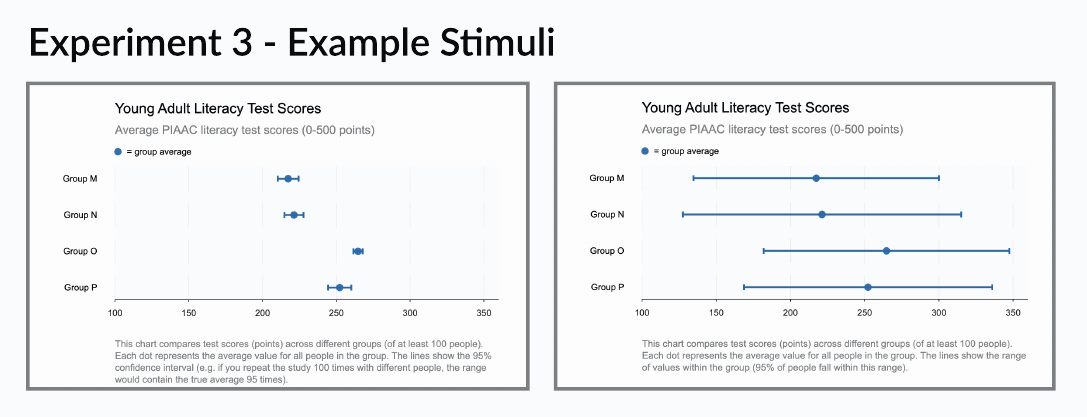}
 \caption{Confidence interval and prediction interval plots used in Experiment 3. }
 \label{Exp3Visualizations}
\vspace{-1em}
\end{figure}


\subsection{Experiment 3: Participants, Design, and Procedure}
We followed the same procedure as Experiment 2 of recruiting participants and the same protocol to filter for spammers and invalid answers. 
We recruited 281 participants, 141 of which saw the confidence interval plot and 140 saw the prediction interval plot.
We followed the same experimental procedure as that in Experiment 2, except that we randomly showed participants either a confidence or a prediction interval plot, as shown in Figure \ref{Exp3Visualizations}. 

\subsection{Experiment 3: Results}
We constructed a mixed-effect linear model to fit participants' ratings of agreement \cite{bates2005fitting} with question type (personal or external) and the type of uncertainty interval visualized (confidence interval or prediction interval), as well as their interactions. 
We additionally included chart topics (literacy scores, household income, and life expectancy), participant race, political orientation, and participant gender as fixed effects.
We used a random intercept term accounting for individual differences and question index as random effects. 

We found a trending significant main effect of uncertainty interval type (F(1, 268.71) = 3.53, p = 0.061, $\eta^2$ = 0.01), and a trending significant effect of question type (F(1, 5.93) = 5.04, p = 0.066, $\eta^2$ = 0.46). Overall participants provided higher rating for the confidence interval plot compared to the prediction interval plot (MD = 3.93, SE = 2.09, d = 0.176), and for the external questions compared to the personal questions (MD = 13.1, SE = 5.83, d = 0.587).
Post-hoc power analysis suggests 100\% power to detect an effect of question type given our sample size. The smaller effect size of the interval type may contribute to us not detecting a significant difference between the two designs, as we are at 38.9\% power at an alpha level of 0.05.

There is no significant interaction between design and question type (F(1, 835.67) = 0.70, p = 0.40). 
However, post-hoc comparisons with Tukey's adjustment on the interaction pairs suggest that, while there is no significant difference between agreement ratings on the external explanations for the two uncertainty interval types (MD = 2.82, p = 0.26), there is a significant difference between agreement ratings on the personal explanations.
See Figure \ref{fig:Overall_Results}.
Specifically, participants who saw the visualizations generated using confidence intervals agreed more with personal explanations of data patterns compared to the participants who saw the visualizations generated using prediction intervals (MD = 5.05, p = 0.04, d = 0.226). 

\begin{figure}[h!]
\centering
   \includegraphics[width = \columnwidth]{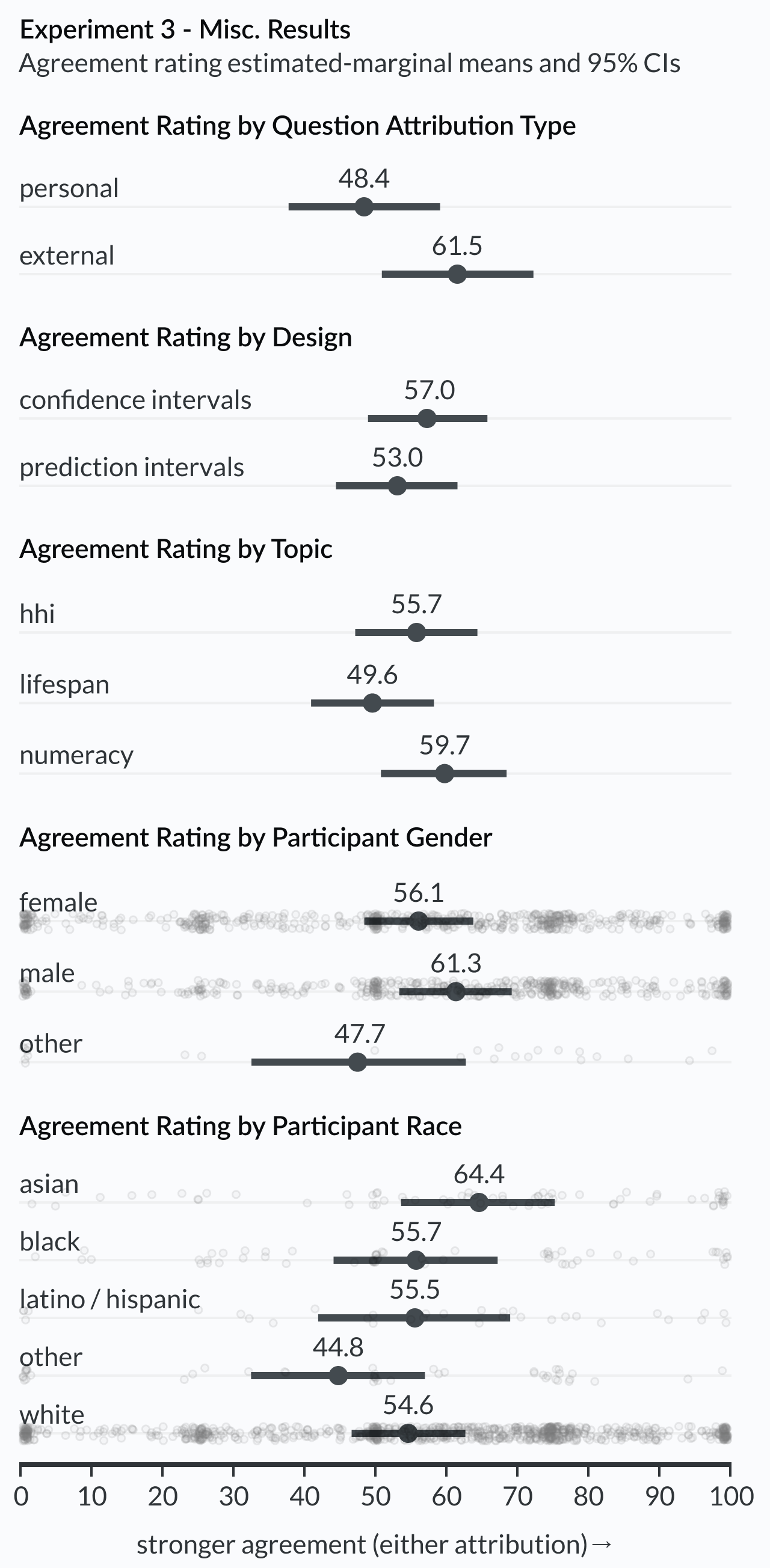}
 \caption{Experiment 3 agreement rating estimated-marginal means and 95\% CIs across factors for both personal and external attributions. First: Ratings by question attribution type. Second: Ratings for chart type. Third: Ratings by topic. Fourth: Ratings by participants' self-identified gender. Fifth: Ratings by participants' self-identified race/ethnicity. Gender and race charts also show jitter plots of individual responses. Results and CIs are based on estimated marginal means (interval overlap should not be used for inference).}
 \label{Exp3TopicGenderRace}
\vspace{-1em}
\end{figure}

\subsubsection{Effects of Topic and Participant Demographics}
We found a significant effect of topic (F(2, 269.48) = 7.24, p < 0.001, $\eta^2$ = 0.051), such that participants agreed the least with both external and personal attributions when reading the visualization depicting the life expectancy topic. See the third panel in Figure \ref{Exp3TopicGenderRace}.

As shown in the bottom panels of Figure \ref{Exp3TopicGenderRace}, there is a significant effect of gender (F(2, 269.03) = 3.78, p = 0.024, $\eta^2$ = 0.027), as well as a trending significant effect of race (F(4, 268.94) = 2.24, p = 0.065, $\eta^2$ = 0.032).
We observed no significant effect of political orientation (F(2, 268.91) = 0.54, p = 0.58).

\subsection{Experiment 3: Discussion}
Consistent with Experiments 1 and 2, personal attribution agreement depended on the chart type and was lower for the chart that visually emphasized outcome variability (the prediction interval). Even though both charts show uncertainty, the chart visualizing outcome uncertainty (prediction intervals, based on the standard deviation of outcomes, showing wider bars) reduced personal attribution agreement compared to the chart visualizing inferential uncertainty of the mean (confidence intervals, based on standard errors, showing smaller bars).


Our results are also consistent with Hofman et al.'s study, showing that confidence intervals exaggerate viewers' judgments about effect size, compared to prediction intervals \cite{hofman2020visualizing}. The authors speculate that confidence intervals give a false impression of low variance for within-group outcomes, nudging viewers toward false dichotomies \cite{besanccon2019continued} (i.e., that all members of one group have superior outcomes to all members of another). Our results support their predictions, further suggesting that confidence intervals can be misread as indicators of outcome dispersion, giving false impressions of intra-group homogeneity, leading to stronger personal attributions, and therefore increased stereotyping.


\begin{figure}[h!]
\centering
 \includegraphics[width = \columnwidth]{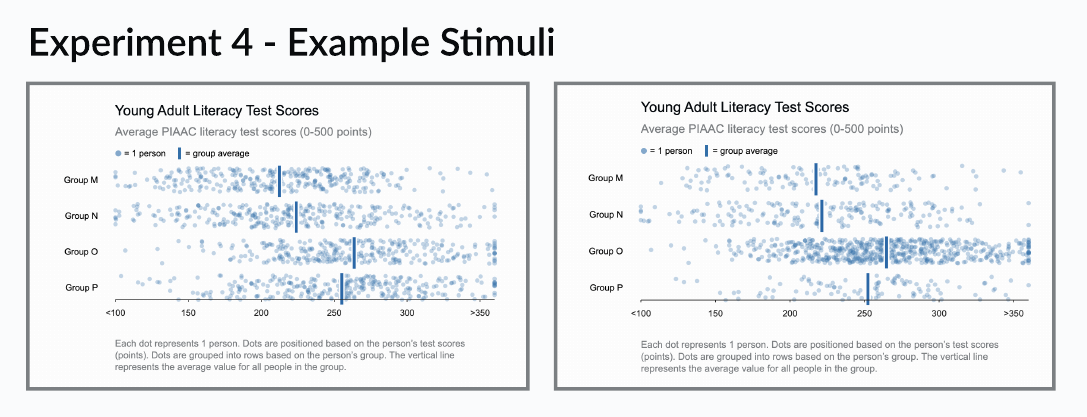}
 \caption{Jitter plot with unequal group sample size and jitter plot with equal group sample size used in Experiment 4.}
 \label{Exp4Visualizations}
\vspace{-1em}
\end{figure}


\section{Experiment 4: Within-Group Sample Size}
In Experiment 1 and Experiment 2, we tested the effect of showing within-group variation with jitter plots on mitigating attribution bias. 
We generated two types of jitter plots: one with equal group sample size (n = 300, Experiment 2), and one with unequal group sample size (Experiment 1). 
While the dot plot with an unequal group sample size more closely resembles real-world data set, it is not yet clear whether the relative number of data points within each group will have an effect on attribution bias (as measured by agreement ratings). In this experiment, we compare participant agreement ratings on external and personal attributions between a jitter plot with an unequal group sample size and a jitter plot with an equal group sample size. 

\subsection{Experiment 4: Participants, Design, and Procedure}
We followed the same recruiting, exclusion protocols, experimental design, and procedure as the previous experiment, except that we randomly showed participants either the jitter plot with equal group sample size (n = 300 dots per group) or unequal group sample size, as shown in Figure \ref{Exp4Visualizations}. 
After applying the exclusion criteria, 266 participants remain, with 132 in the equal sample size group and 134 in the unequal sample size group.

\subsection{Experiment 4: Results}
We constructed a mixed-effect linear model to fit participants' ratings of agreement \cite{bates2005fitting} with question type (personal or external) and the sample size choice (even sample size across groups or uneven sample size across groups), as well as their interactions.
Same with our previous models, we included chart topics (literacy scores, household income, and life expectancy), participant race, political orientation, and participant gender as fixed effects.
For random effects, we used a random intercept term accounting for individual differences and questions. 

We found no significant main effect of sample size choice (F(1, 253.50) = 0.95, p = 0.629).
But we did find a significant main effect of question type (F(1, 5.93) = 7.14, p = 0.037, $eta^2$ = 0.55), such that participants agreed more with the external explanations compared to the personal explanations (MD = 15.0, SE = 5.63, d = 0.654).
There is no significant interaction effect between question type and chart design (F(1, 790.48) = 1.03, p = 0.31). 
Post-hoc comparisons with Tukey's adjustment on the interaction pairs further suggest that the agreement ratings for neither personal nor external explanations differed depending on the sample size choice. 
As shown in Figure \ref{fig:Overall_Results}, participants that viewed jitter plots with an unequal sample size across groups agreed equally as much as participants that viewed jitter plots with an equal sample size across groups. 

\begin{figure}[h!]
\centering
   \includegraphics[width = \columnwidth]{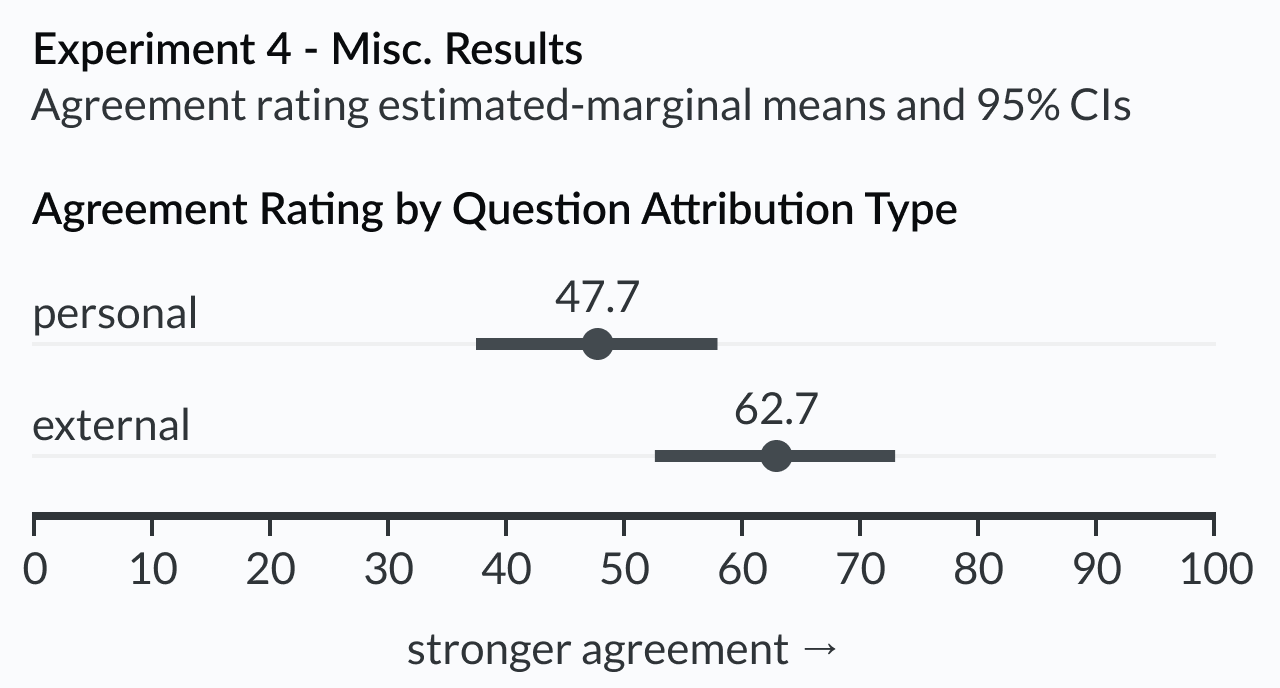}
 \caption{Experiment 4 agreement rating estimated-marginal means and 95\% CIs across factors for both personal and external attributions, by question attribution type. Results and CIs are based on estimated marginal means (interval overlap should not be used for inference).}
 \label{Exp4TopicGenderRace}
\vspace{-1em}
\end{figure}

\subsubsection{Effects of Topic and Participant Demographics}
We did not find a significant effect of topic (F(2, 253.94) = 2.08, p = 0.13). Participants were equally likely to agree with personal and external explanations across all three topics. 
We also observed no evidence of a significant effect of gender (F(2, 254.01) = 0.85, p = 0.43), nor of political orientation (F(2, 253.78) = 1.22, p = 0.30), nor of race (F(4, 254.16) = 0.74, p = 0.56).

\subsection{Experiment 4: Discussion}

We observed no difference in attribution agreement between the two conditions, suggesting that findings from Experiment 2 (where we used a jitter plot with equal sample sizes between groups) would generalize to more realistic datasets with uneven group sample sizes. 

However, the effect of relative group sample size on attribution bias (or on visualization interpretation in general) remains an interesting topic. Existing work in human cognition suggests that the number of data points can influence viewer perception and interpretation of data \cite{cowan2001magical, hollingworth2008understanding, zhang2008discrete}. For example, adding more data points to a visualization adds substantial noise to how readers mentally represent the visualization and subsequently recall the visualization \cite{mccoleman2021rethinking}, and the amount of data disclosed in a visualization can impact viewer trust \cite{xiong2019examining}. People also perceive the correlations of a smaller sample size as more extreme \cite{kareev1997through}. There is a also general cognitive tendency for people to use the number of people in a group to make group-based social judgments \cite{price2006effect, voci2008majority}. When the group sample size is large, people become less likely to perceive the members of the group as individuals and more likely to treat the group as a whole \cite{kerr1989illusions} (which is usually used to explain the bystander effect \cite{fischer2011bystander}). In light of this work, we speculate that viewers might be less likely to consider individual differences when the group size is large (e.g., group D in the chart). They might attribute disparities between groups with large sample sizes as something fundamentally different between the groups, with diminished considerations for within-group variations, and fall victim to attribution bias. 

On the other hand, it is possible that the number of data points within a group does not have an effect. For example, in social psychology examining the effect of social conformity, the number of members within a group has little effect on conformity behaviors, rather, the total number of distinct groups matters more \cite{wilder1977perception}. In the workplace, people also do not perceive small groups to be significantly different from large groups in terms of productivity and commitment \cite{ogungbamila2010effects}. So if these results generalize to interpretations of data visualizations, we would expect the number of data points within each group to have less effect on attribution bias. 

While the current Experiment 4 was not designed to test for the effect of relative group sample size, this is an interesting and under-explored topic for future research. We offer more discussion and suggestions with regards to this topic in Section \ref{limitsFuture}.

\section{General Discussion}
\label{Discussion}

In Experiment 1, participants saw either a bar chart (sans variability indicators) or a jitter plot. We found that bar charts increased personal attribution agreement compared to jitter plots. This result supports \textbf{Hypothesis No.1} and \textbf{Hypothesis No.2}. 


In Experiment 2, to better isolate the influential visualization factors, participants saw one of three chart types: a dot plot (sans variability indicators), a jitter plot, or a prediction interval. We found that, consistent with before, the chart masking intra-group variability (the plain dot plot) elicited stronger personal attribution. This result supports \textbf{Hypothesis No.1} and \textbf{Hypothesis No.2}. 

In Experiment 3, to understand the effect of visualizing different types of uncertainty, participants saw either confidence intervals (representing inferential uncertainty) or prediction intervals (representing outcome uncertainty). We found that confidence intervals elicited stronger personal attribution. This result corroborates existing findings \cite{belia2005researchers} and supports \textbf{Hypothesis No.3}.

In Experiment 4, we compared jitter plots with equal numbers of within-group data points to plots with unequal numbers of within-group data points. We found no difference in participants' attributions. This result does not support \textbf{Hypothesis No.4}.


Overall, these experiments suggest that readers can misinterpret charts about inter-group outcome disparities in ways that reinforce stereotypes and that visualizations emphasizing outcome variability can reduce this effect. The charts we showed participants only showed data about outcomes, not \emph{why} the outcomes occurred; however a substantial number of participants misinterpreted the charts as evidence supporting personal attributions (i.e., that outcome differences were caused by differences in personal characteristics of each group).

An even larger subset of participants misinterpreted the charts as evidence supporting external attributions (i.e., that outcome differences were due to external forces outside the groups' control). While these conclusions were also unsupported by the stimuli charts, this shows participants were more willing to take an empathetic view of outcome disparities, which is perhaps a win for humanity, if not rationality.

We found that participants' tendencies toward personal attributions can be influenced by visualization design choices. Specifically, visualizations that emphasize within-group outcome variability (i.e., jitter plots, prediction intervals) reduced agreement with personal attributions relative to visualizations that hide or minimize it (i.e., bar charts, dot plots, confidence intervals). Visualization design did not significantly influence external attributions. 

The results support arguments that conventional visualization approaches may be harmful in the context of racial equity \cite{blakely2021presenting, schwabish2021no} and suggest that when visualizing data about people, particularly members of minoritized communities, designers have a duty of care beyond accurately representing the facts of the data.
The results also underscore the rhetorical nature of visualizations by demonstrating a novel way that framing can impact viewers’ interpretations of the data \cite{hullman2011rhetoric}.
\newline
\vspace{-2mm}

\noindent \textbf{Design Implications:} When visualizing data about people, especially those in minoritized groups, designers should consider the social-cognitive impacts of their design choices. Conventional approaches (e.g., bar charts) that are safe in other contexts may introduce unexpected harms. 
Our work also suggests that group averages alone are insufficient for responsibly visualizing social outcome disparities. To reduce social-cognitive biases that lead to harmful stereotyping, visualizations should show the diversity of outcomes within groups, not just the differences between them.

\section{Limitation and Future Directions}
\label{limitsFuture}
As an initial investigation of how visualization design can impact reader perceptions of people and stereotypes, we discuss several limitations and suggest a path forward for future experiments.

\noindent \textbf{Visualization effects on external attributions:} Interestingly, we found no significant differences between visualization design conditions with respect to external attributions. This may be related to the experiment design, where the questions themselves prime users with plausible explanations for the visualized disparities (and 50\% of the questions implied external explanations). Asking for open-ended responses may produce different results. The results do suggest, though, that when viewers are reminded of possible external explanations, or they are already familiar with them (i.e., restaurant worker pay), then conventional approaches do not seem to diminish agreement with external attributions. Under these limited conditions the results suggest no significant ``distraction'' effects between design conditions. Future work can leverage other methods to further investigate this effect. 
\newline
\vspace{-2mm}

\begin{table}[h!]
    \setlength\extrarowheight{2pt}
    \centering
\begin{tabular}{l|l|l|l|l}
          & Exp 1                                                            & Exp 2                              & Exp 3                                     & Exp 4    \\ \hline
Race      & p = 0.44                                                         & \cellcolor[HTML]{DAE8FC}p = 0.0078 & \cellcolor[HTML]{F8FAFE}p = 0.065         & p = 0.56 \\ \hline
Political & \cellcolor[HTML]{C2DBFE}{\color[HTML]{333333} p \textless 0.001} & p = 0.11                           & \cellcolor[HTML]{F8FAFE}p = 0.058         & p = 0.30 \\ \hline
Gender    & -                                                                & \cellcolor[HTML]{ECF4FF}p = 0.027  & \cellcolor[HTML]{ECF4FF}p = 0.024         & p = 0.43 \\ \hline
Topic     & -                                                                & \cellcolor[HTML]{DAE8FC}p = 0.0028 & \cellcolor[HTML]{C2DBFE}p \textless 0.001 & p = 0.13
    \end{tabular}
    \vspace{2px}
    \caption{Summary of the effect of demographic and topic factors. Darker hue represents more statistically significant differences.}
    \label{demoTable}
\end{table}

\noindent \textbf{Individual Differences and Effect of Topic:} We observed some effects of participant gender, race, and political orientation in our study, as well as an effect of the visualization topic. 
See Table \ref{demoTable} for a comparison of their effects between experiments. 
Note that the present experiment was not designed nor powered to identify potential individual differences in attribution bias, such as the effect of political orientation. But future work can do larger-scaled studies to more deeply investigate the effects of demographic factors, and model bias-driven misinterpretation of visualizations, such as accounting for the participants' individual prior beliefs on the topics depicted or exploring potential interaction effects between political orientation and external versus internal attributions during data interpretation.
\newline
\vspace{-2mm}

\noindent \textbf{Anthropomorphic Visualizations:} We found that participants tend to attribute outcome disparities to personal factors more when reading charts that hide within-group variability, and less so when reading charts that show within-group variability. While prediction intervals and jitter plots both show within-group outcome dispersion, it is possible that jitter plot perceptions are influenced by visualization anthropomorphism, such that the individual points plotted motivate readers to think about the individual people the data represents \cite{boy2017showing, bigTriangle, ZerAviv2015Unemmpathetic}. To the extent that anthropomorphism influences empathy, and empathy influences attribution errors, this may introduce another dynamic worth exploring. Future experiments can investigate this by comparing charts that show aggregated means only, charts that show individual participant's data, or charts that represent individual participants with anthropomorphic icons, such as in \cite{weePeople, burns2021designing}.
\newline
\vspace{-2mm}

\noindent \textbf{Other Measures:} Given social-desirability biases, measuring stereotyping requires indirect, sometimes creative approaches. We used agreement ratings with statements describing personal and external attributions, but surveys like this can be subject to self-reporting biases \cite{adams1999evidence}. Our participants reacted differently depending on whether racial groups were explicitly labeled on the chart or not. This might suggest that the presence of racial information intuitively triggered social-desirability biases, causing participants to artificially suppress personal attributions in their self-reporting. Future studies can consider using other methodologies such as hypothetical scenarios \cite{strickland2022behavioral} or test for implicit biases \cite{brosch2013implicit} to account for this effect. The present study also only tested bar charts and versions of dot plots (e.g., with and without intervals, with jitters). Future work can explore the effect of other visualization designs such as color choices or other chart types. 
\newline
\vspace{-2mm}

\noindent \textbf{Sample Size and Blocking:} In the current study the number of participants in each experimental condition was not identical. We chose to keep extra participants we sampled and conducted post-hoc power analysis to compare its effect, but future work should leverage the blocking features in online crowdsourcing platforms to ensure the sample size in each condition match each other or additionally randomly select to exclude participants from conditions to match the sample sizes while keeping with the power analysis.
\newline
\vspace{-2mm}

\noindent \textbf{Jitter and Visual Differences:} We used the same set of samples to generate jitter plots and only tested four sets of means using four topics. The dataset used was based on real data. 
We kept the dataset consistent throughout the experiment to maximize control, so we could identify the exact effect of attribution bias without the number of conditions exploding \cite{wall2022vishikers}. However, as demonstrated in \cite{xiong2022investigating} and \cite{jardine2019perceptual}, the arrangement of jittered patterns and the perceptual proxies of the data (driven by the shape of different means) can also influence viewer takeaways from the visualization. Future experiments can further test out a range of effects on attribution bias by varying the visual differences in the underlying dataset, including changing the jittered patterns, group means, and within-group data variability.
\newline
\vspace{-2mm}

\acknowledgments{
We are grateful for the participants who made this study possible. We thank Pieta Blakely, Jon Schwabish, Alice Feng, Mary Aviles, Melissa Kovacs, David Napoli, Kevin Ford, Ken Choi, Robert Kosara, Steve Franconeri, Steve Haroz, and Akira Wada for their inspiration, advice, and input. We also thank our reviewers for their helpful feedback. 
}

\newpage
\bibliographystyle{abbrv}
\bibliography{reference}

\begin{thebibliography}{10}

\bibitem{adams1999evidence}
A.~S. Adams, S.~B. Soumerai, J.~Lomas, and D.~Ross-Degnan.
\newblock Evidence of self-report bias in assessing adherence to guidelines.
\newblock {\em International Journal for Quality in Health Care},
  11(3):187--192, 1999.

\bibitem{agley2022quality}
J.~Agley, Y.~Xiao, R.~Nolan, and L.~Golzarri-Arroyo.
\newblock Quality control questions on amazon’s mechanical turk (mturk): A
  randomized trial of impact on the usaudit, phq-9, and gad-7.
\newblock {\em Behavior research methods}, 54(2):885--897, 2022.

\bibitem{allison1985group}
S.~T. Allison and D.~M. Messick.
\newblock The group attribution error.
\newblock {\em Journal of Experimental Social Psychology}, 21(6):563--579,
  1985.

\bibitem{andrade2011excessive}
E.~B. Andrade.
\newblock Excessive confidence in visually-based estimates.
\newblock {\em Organizational Behavior and Human Decision Processes},
  116(2):252--261, 2011.

\bibitem{bates2005fitting}
D.~Bates et~al.
\newblock Fitting linear mixed models in r.
\newblock {\em R news}, 5(1):27--30, 2005.

\bibitem{belia2005researchers}
S.~Belia, F.~Fidler, J.~Williams, and G.~Cumming.
\newblock Researchers misunderstand confidence intervals and standard error
  bars.
\newblock {\em Psychological methods}, 10(4):389, 2005.

\bibitem{besanccon2019continued}
L.~Besan{\c{c}}on and P.~Dragicevic.
\newblock The continued prevalence of dichotomous inferences at chi.
\newblock In {\em Extended Abstracts of the 2019 CHI Conference on Human
  Factors in Computing Systems}, pages 1--11, 2019.

\bibitem{blakely2021presenting}
P.~Blakely.
\newblock Presenting data for a targeted universalist approach, 2021.

\bibitem{blanco2013interactive}
F.~Blanco, H.~Matute, and M.~A. Vadillo.
\newblock Interactive effects of the probability of the cue and the probability
  of the outcome on the overestimation of null contingency.
\newblock {\em Learning \& Behavior}, 41(4):333--340, 2013.

\bibitem{boy2017showing}
J.~Boy, A.~V. Pandey, J.~Emerson, M.~Satterthwaite, O.~Nov, and E.~Bertini.
\newblock Showing people behind data: Does anthropomorphizing visualizations
  elicit more empathy for human rights data?
\newblock In {\em Proceedings of the 2017 CHI Conference on Human Factors in
  Computing Systems}, pages 5462--5474, 2017.

\bibitem{brosch2013implicit}
T.~Brosch, E.~Bar-David, and E.~A. Phelps.
\newblock Implicit race bias decreases the similarity of neural representations
  of black and white faces.
\newblock {\em Psychological science}, 24(2):160--166, 2013.

\bibitem{burns2021designing}
A.~Burns, C.~Xiong, S.~Franconeri, A.~Cairo, and N.~Mahyar.
\newblock Designing with pictographs: Envision topics without sacrificing
  understanding.
\newblock {\em IEEE transactions on visualization and computer graphics}, 2021.

\bibitem{cleveland1986experiment}
W.~S. Cleveland and R.~McGill.
\newblock An experiment in graphical perception.
\newblock {\em International Journal of Man-Machine Studies}, 25(5):491--500,
  1986.

\bibitem{correll2014error}
M.~Correll and M.~Gleicher.
\newblock Error bars considered harmful: Exploring alternate encodings for mean
  and error.
\newblock {\em IEEE transactions on visualization and computer graphics},
  20(12):2142--2151, 2014.

\bibitem{cowan2001magical}
N.~Cowan.
\newblock The magical number 4 in short-term memory: A reconsideration of
  mental storage capacity.
\newblock {\em Behavioral and brain sciences}, 24(1):87--114, 2001.

\bibitem{cox2013visualizing}
J.~Cox, D.~House, and M.~Lindell.
\newblock Visualizing uncertainty in predicted hurricane tracks.
\newblock {\em International Journal for Uncertainty Quantification}, 3(2),
  2013.

\bibitem{davis2019deficit}
L.~P. Davis and S.~D. Museus.
\newblock What is deficit thinking? an analysis of conceptualizations of
  deficit thinking and implications for scholarly research.
\newblock {\em NCID Currents}, 1(1), 2019.

\bibitem{stangor2014biases}
D.C.Stangor.
\newblock Biases on attribution, 2014.

\bibitem{stangor2014reducing}
D.C.Stangor.
\newblock Reducing discrimination, 2014.

\bibitem{stangor2014social}
D.C.Stangor.
\newblock Social categorization and stereotyping, 2014.

\bibitem{d2020data}
C.~D'ignazio and L.~F. Klein.
\newblock {\em Data feminism}.
\newblock MIT press, 2020.

\bibitem{fischer2011bystander}
P.~Fischer, J.~I. Krueger, T.~Greitemeyer, C.~Vogrincic, A.~Kastenm{\"u}ller,
  D.~Frey, M.~Heene, M.~Wicher, and M.~Kainbacher.
\newblock The bystander-effect: a meta-analytic review on bystander
  intervention in dangerous and non-dangerous emergencies.
\newblock {\em Psychological bulletin}, 137(4):517, 2011.

\bibitem{goffman1978presentation}
E.~Goffman et~al.
\newblock {\em The presentation of self in everyday life}, volume~21.
\newblock Harmondsworth London, 1978.

\bibitem{grimm2010social}
P.~Grimm.
\newblock Social desirability bias.
\newblock {\em Wiley international encyclopedia of marketing}, 2010.

\bibitem{hamill1980insensitivity}
R.~Hamill, T.~D. Wilson, and R.~E. Nisbett.
\newblock Insensitivity to sample bias: Generalizing from atypical cases.
\newblock {\em Journal of Personality and Social Psychology}, 39(4):578, 1980.

\bibitem{weePeople}
J.~Harris.
\newblock Connecting with the dots.
\newblock 2015.

\bibitem{hofman2020visualizing}
J.~M. Hofman, D.~G. Goldstein, and J.~Hullman.
\newblock How visualizing inferential uncertainty can mislead readers about
  treatment effects in scientific results.
\newblock In {\em Proceedings of the 2020 chi conference on human factors in
  computing systems}, pages 1--12, 2020.

\bibitem{hollingworth2008understanding}
A.~Hollingworth, A.~M. Richard, and S.~J. Luck.
\newblock Understanding the function of visual short-term memory: transsaccadic
  memory, object correspondence, and gaze correction.
\newblock {\em Journal of Experimental Psychology: General}, 137(1):163, 2008.

\bibitem{hopkins2009partisan}
D.~J. Hopkins.
\newblock Partisan reinforcement and the poor: The impact of context on
  explanations for poverty.
\newblock {\em Social Science Quarterly}, 90(3):744--764, 2009.

\bibitem{hullman2019authors}
J.~Hullman.
\newblock Why authors don't visualize uncertainty.
\newblock {\em IEEE transactions on visualization and computer graphics},
  26(1):130--139, 2019.

\bibitem{hullman2011rhetoric}
J.~Hullman and N.~Diakopoulos.
\newblock Visualization rhetoric: Framing effects in narrative visualization.
\newblock {\em IEEE Transactions on Visualization and Computer Graphics},
  17(12):2231--2240, 2011.

\bibitem{hullman2015hypothetical}
J.~Hullman, P.~Resnick, and E.~Adar.
\newblock Hypothetical outcome plots outperform error bars and violin plots for
  inferences about reliability of variable ordering.
\newblock {\em PloS one}, 10(11):e0142444, 2015.

\bibitem{jardine2019perceptual}
N.~Jardine, B.~D. Ondov, N.~Elmqvist, and S.~Franconeri.
\newblock The perceptual proxies of visual comparison.
\newblock {\em IEEE transactions on visualization and computer graphics},
  26(1):1012--1021, 2019.

\bibitem{kahneman2011thinking}
D.~Kahneman.
\newblock {\em Thinking, fast and slow}.
\newblock Macmillan, 2011.

\bibitem{kahneman1982judgment}
D.~Kahneman, S.~P. Slovic, P.~Slovic, and A.~Tversky.
\newblock {\em Judgment under uncertainty: Heuristics and biases}.
\newblock Cambridge university press, 1982.

\bibitem{kale2020visual}
A.~Kale, M.~Kay, and J.~Hullman.
\newblock Visual reasoning strategies for effect size judgments and decisions.
\newblock {\em IEEE transactions on visualization and computer graphics},
  27(2):272--282, 2020.

\bibitem{kale2018hypothetical}
A.~Kale, F.~Nguyen, M.~Kay, and J.~Hullman.
\newblock Hypothetical outcome plots help untrained observers judge trends in
  ambiguous data.
\newblock {\em IEEE transactions on visualization and computer graphics},
  25(1):892--902, 2018.

\bibitem{kareev1997through}
Y.~Kareev, I.~Lieberman, and M.~Lev.
\newblock Through a narrow window: Sample size and the perception of
  correlation.
\newblock {\em Journal of Experimental Psychology: General}, 126(3):278, 1997.

\bibitem{kay2016ish}
M.~Kay, T.~Kola, J.~R. Hullman, and S.~A. Munson.
\newblock When (ish) is my bus? user-centered visualizations of uncertainty in
  everyday, mobile predictive systems.
\newblock In {\em Proceedings of the 2016 chi conference on human factors in
  computing systems}, pages 5092--5103, 2016.

\bibitem{kennedy2020shape}
R.~Kennedy, S.~Clifford, T.~Burleigh, P.~D. Waggoner, R.~Jewell, and N.~J.
  Winter.
\newblock The shape of and solutions to the mturk quality crisis.
\newblock {\em Political Science Research and Methods}, 8(4):614--629, 2020.

\bibitem{kerns2021bar}
S.~Kerns and J.~Wilmer.
\newblock The bar-tip limit error: a common, qualitative misinterpretation of
  bar graphs of means revealed by the ddog method.
\newblock {\em Journal of Vision}, 21(9):2602--2602, 2021.

\bibitem{kerns2021two}
S.~H. Kerns and J.~B. Wilmer.
\newblock Two graphs walk into a bar: Readout-based measurement reveals the
  bar-tip limit error, a common, categorical misinterpretation of mean bar
  graphs.
\newblock {\em Journal of vision}, 21(12):17--17, 2021.

\bibitem{kerr1989illusions}
N.~L. Kerr.
\newblock Illusions of efficacy: The effects of group size on perceived
  efficacy in social dilemmas.
\newblock {\em Journal of experimental social psychology}, 25(4):287--313,
  1989.

\bibitem{krzywinski2013errorbars}
M.~Krzywinski and N.~Altman.
\newblock Points of significance: Error bars.
\newblock {\em Nature methods}, 10:921--2, 09 2013.

\bibitem{lan2022negative}
X.~Lan, Y.~Wu, Y.~Shi, Q.~Chen, and N.~Cao.
\newblock Negative emotions, positive outcomes? exploring the communication of
  negativity in serious data stories.
\newblock In {\em CHI Conference on Human Factors in Computing Systems}, pages
  1--14, 2022.

\bibitem{lerner1965evaluation}
M.~J. Lerner.
\newblock Evaluation of performance as a function of performer's reward and
  attractiveness.
\newblock {\em Journal of personality and social psychology}, 1(4):355, 1965.

\bibitem{linville1980polarized}
P.~W. Linville and E.~E. Jones.
\newblock Polarized appraisals of out-group members.
\newblock {\em Journal of personality and social psychology}, 38(5):689, 1980.

\bibitem{manski2020lure}
C.~F. Manski.
\newblock The lure of incredible certitude.
\newblock {\em Economics \& Philosophy}, 36(2):216--245, 2020.

\bibitem{matute2015illusions}
H.~Matute, F.~Blanco, I.~Yarritu, M.~D{\'\i}az-Lago, M.~A. Vadillo, and
  I.~Barberia.
\newblock Illusions of causality: how they bias our everyday thinking and how
  they could be reduced.
\newblock {\em Frontiers in psychology}, 6:888, 2015.

\bibitem{bigTriangle}
S.~McCloud.
\newblock The big triangle.
\newblock 1999.

\bibitem{mccoleman2021rethinking}
C.~M. McColeman, F.~Yang, T.~F. Brady, and S.~Franconeri.
\newblock Rethinking the ranks of visual channels.
\newblock {\em IEEE Transactions on Visualization and Computer Graphics},
  28(1):707--717, 2021.

\bibitem{newman2012bar}
G.~E. Newman and B.~J. Scholl.
\newblock Bar graphs depicting averages are perceptually misinterpreted: The
  within-the-bar bias.
\newblock {\em Psychonomic bulletin \& review}, 19(4):601--607, 2012.

\bibitem{nguyen2020exploring}
F.~Nguyen, X.~Qiao, J.~Heer, and J.~Hullman.
\newblock Exploring the effects of aggregation choices on untrained
  visualization users' generalizations from data.
\newblock In {\em Computer graphics forum}, volume~39, pages 33--48. Wiley
  Online Library, 2020.

\bibitem{oakes1994stereotyping}
P.~J. Oakes, S.~A. Haslam, and J.~C. Turner.
\newblock {\em Stereotyping and social reality.}
\newblock Blackwell Publishing, 1994.

\bibitem{waiters2021waitresses}
U.~B. of~Labor~Statistics.
\newblock Occupational employment and wage statistics, 2020.

\bibitem{economic2021table_wage}
U.~B. of~Labor~Statistics.
\newblock Table 3. median usual weekly earnings of full-time wage and salary
  workers by age, race, hispanic or latino ethnicity, and sex, fourth quarter
  2021 averages, not seasonally adjusted, 2021.

\bibitem{ogungbamila2010effects}
B.~Ogungbamila, A.~Ogungbamila, and G.~Agboola~Adetula.
\newblock Effects of team size and work team perception on workplace
  commitment: Evidence from 23 production teams.
\newblock {\em Small group research}, 41(6):725--745, 2010.

\bibitem{osborne2015latent}
D.~Osborne and B.~Weiner.
\newblock A latent profile analysis of attributions for poverty: Identifying
  response patterns underlying people’s willingness to help the poor.
\newblock {\em Personality and Individual Differences}, 85:149--154, 2015.

\bibitem{padilla2017effects}
L.~M. Padilla, I.~T. Ruginski, and S.~H. Creem-Regehr.
\newblock Effects of ensemble and summary displays on interpretations of
  geospatial uncertainty data.
\newblock {\em Cognitive research: principles and implications}, 2(1):1--16,
  2017.

\bibitem{park1982perception}
B.~Park and M.~Rothbart.
\newblock Perception of out-group homogeneity and levels of social
  categorization: Memory for the subordinate attributes of in-group and
  out-group members.
\newblock {\em Journal of Personality and Social Psychology}, 42(6):1051, 1982.

\bibitem{pettigrew2006meta}
T.~F. Pettigrew and L.~R. Tropp.
\newblock A meta-analytic test of intergroup contact theory.
\newblock {\em Journal of personality and social psychology}, 90(5):751, 2006.

\bibitem{price2006effect}
P.~C. Price, A.~R. Smith, and H.~C. Lench.
\newblock The effect of target group size on risk judgments and comparative
  optimism: the more, the riskier.
\newblock {\em Journal of Personality and Social Psychology}, 90(3):382, 2006.

\bibitem{ruginski2016non}
I.~T. Ruginski, A.~P. Boone, L.~M. Padilla, L.~Liu, N.~Heydari, H.~S. Kramer,
  M.~Hegarty, W.~B. Thompson, D.~H. House, and S.~H. Creem-Regehr.
\newblock Non-expert interpretations of hurricane forecast uncertainty
  visualizations.
\newblock {\em Spatial Cognition \& Computation}, 16(2):154--172, 2016.

\bibitem{samson2005seeing}
D.~Samson, I.~A. Apperly, U.~Kathirgamanathan, and G.~W. Humphreys.
\newblock Seeing it my way: a case of a selective deficit in inhibiting
  self-perspective.
\newblock {\em Brain}, 128(5):1102--1111, 2005.

\bibitem{schwabish2021no}
J.~Schwabish and A.~Feng.
\newblock Do no harm guide: Applying equity awareness in data visualization.
\newblock {\em https://www. urban.
  org/research/publication/do-no-harm-guide-applying-equity-awareness-data-visualization},
  2021.

\bibitem{sears2005over}
D.~O. Sears and P.~J. Henry.
\newblock Over thirty years later: A contemporary look at symbolic racism.
\newblock {\em Advances in experimental social psychology}, 37(1):95--125,
  2005.

\bibitem{soyer2012illusion}
E.~Soyer and R.~M. Hogarth.
\newblock The illusion of predictability: How regression statistics mislead
  experts.
\newblock {\em International Journal of Forecasting}, 28(3):695--711, 2012.

\bibitem{srinivasan2018s}
A.~Srinivasan, M.~Brehmer, B.~Lee, and S.~M. Drucker.
\newblock What's the difference? evaluating variations of multi-series bar
  charts for visual comparison tasks.
\newblock In {\em Proceedings of the 2018 CHI Conference on Human Factors in
  Computing Systems}, pages 1--12, 2018.

\bibitem{stark2019impact}
T.~H. Stark, F.~M. van Maaren, J.~A. Krosnick, and G.~Sood.
\newblock The impact of social desirability pressures on whites’ endorsement
  of racial stereotypes: A comparison between oral and acasi reports in a
  national survey.
\newblock {\em Sociological Methods \& Research}, page 0049124119875959, 2019.

\bibitem{streit2014bar}
M.~Streit and N.~Gehlenborg.
\newblock Bar charts and box plots: creating a simple yet effective plot
  requires an understanding of data and tasks.
\newblock {\em Nature methods}, 11(2):117--118, 2014.

\bibitem{strickland2022behavioral}
J.~C. Strickland, D.~D. Reed, S.~R. Hursh, L.~P. Schwartz, R.~N. Foster, B.~W.
  Gelino, R.~S. LeComte, F.~S. Oda, A.~R. Salzer, T.~D. Schneider, et~al.
\newblock Behavioral economic methods to inform infectious disease response:
  Prevention, testing, and vaccination in the covid-19 pandemic.
\newblock {\em PloS one}, 17(1):e0258828, 2022.

\bibitem{szafir2018good}
D.~A. Szafir.
\newblock The good, the bad, and the biased: Five ways visualizations can
  mislead (and how to fix them).
\newblock {\em interactions}, 25(4):26--33, 2018.

\bibitem{voci2008majority}
A.~Voci, M.~Hewstone, R.~J. Crisp, and M.~Rubin.
\newblock Majority, minority, and parity: Effects of gender and group size on
  perceived group variability.
\newblock {\em Social Psychology Quarterly}, 71(2):114--142, 2008.

\bibitem{wall2022vishikers}
E.~Wall, C.~Xiong, and Y.-S. Kim.
\newblock Vishikers’ guide to evaluation: Competing considerations in study
  design.
\newblock {\em IEEE Computer Graphics and Applications}, 42(3):29--38, 2022.

\bibitem{weiner2011attributional}
B.~Weiner, D.~Osborne, and U.~Rudolph.
\newblock An attributional analysis of reactions to poverty: The political
  ideology of the giver and the perceived morality of the receiver.
\newblock {\em Personality and Social Psychology Review}, 15(2):199--213, 2011.

\bibitem{wilder1977perception}
D.~A. Wilder.
\newblock Perception of groups, size of opposition, and social influence.
\newblock {\em Journal of Experimental Social Psychology}, 13(3):253--268,
  1977.

\bibitem{xiong2019biased}
C.~Xiong, C.~R. Ceja, C.~J. Ludwig, and S.~Franconeri.
\newblock Biased average position estimates in line and bar graphs:
  Underestimation, overestimation, and perceptual pull.
\newblock {\em IEEE transactions on visualization and computer graphics},
  26(1):301--310, 2019.

\bibitem{xiong2019examining}
C.~Xiong, L.~Padilla, K.~Grayson, and S.~Franconeri.
\newblock Examining the components of trust in map-based visualizations.
\newblock 2019.

\bibitem{xiong2022investigating}
C.~Xiong, A.~Sarvghad, D.~G. Goldstein, J.~M. Hofman, and {\c{C}}.~Demiralp.
\newblock Investigating perceptual biases in icon arrays.
\newblock In {\em CHI Conference on Human Factors in Computing Systems}, pages
  1--12, 2022.

\bibitem{xiong2021visual}
C.~Xiong, V.~Setlur, B.~Bach, E.~Koh, K.~Lin, and S.~Franconeri.
\newblock Visual arrangements of bar charts influence comparisons in viewer
  takeaways.
\newblock {\em IEEE Transactions on Visualization and Computer Graphics},
  28(1):955--965, 2021.

\bibitem{xiong2019illusion}
C.~Xiong, J.~Shapiro, J.~Hullman, and S.~Franconeri.
\newblock Illusion of causality in visualized data.
\newblock {\em IEEE transactions on visualization and computer graphics},
  26(1):853--862, 2019.

\bibitem{ZerAviv2015Unemmpathetic}
M.~Zer-Aviv.
\newblock Dataviz—the unempathetic art.
\newblock 2015.

\bibitem{zhang2008discrete}
W.~Zhang and S.~J. Luck.
\newblock Discrete fixed-resolution representations in visual working memory.
\newblock {\em Nature}, 453(7192):233--235, 2008.

\end{thebibliography}

\end{document}